\documentclass{article}
\usepackage[utf8]{inputenc}
\usepackage{longtable}
\usepackage{booktabs}
\usepackage{graphicx}
\usepackage{subcaption}
\usepackage{appendix}
\usepackage{amssymb}
\usepackage{amsmath} 
\usepackage{comment}
\usepackage{hyperref}
\usepackage{listings}
\AtBeginDocument{}
\usepackage[capitalise,noabbrev]{cleveref}
\usepackage[english]{babel}
\usepackage[nottoc,numbib]{tocbibind}
\usepackage{color}
\usepackage{csquotes}
\usepackage{tablefootnote}
\usepackage{pifont}
\usepackage{framed}
\usepackage{bsymb}
\usepackage{biblatex}[backend=biber, maxnames=30, style = unsrt]
\usepackage{adjustbox}

\begin{document}

\newcommand{\btoprogram}{{\sc B2Program}}
\newcommand{\prob}{{\sc ProB}}
\newcommand{\simb}{{\sc SimB}}

\newcommand{\cmark}{\ding{51}}%
\newcommand{\xmark}{\ding{55}}%
\newcommand{\questionmark}{\textbf{?}}%

\begin{titlepage}
   \centering

    \hfil\includegraphics[width=8cm]{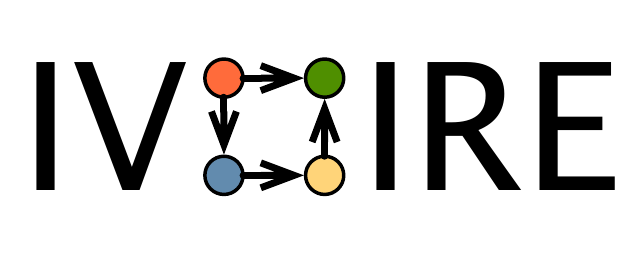}\newline
    \hfil\hfil\includegraphics[width=5cm]{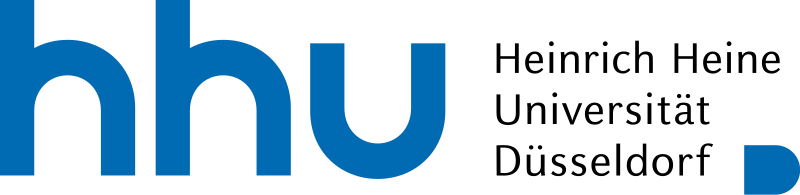}
    \hfil\hfil\hfil\hfil\includegraphics[width=4cm]{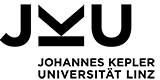}\newline
    \par\vspace*{4\baselineskip}
    {\huge \bfseries IVOIRE -  Deliverable D 1.2}\\[1.5cm]
\textsc{\Large{\today}}\\[6.0cm]

Sebastian Stock, Fabian Vu, David Gele\ss{}us, Atif Mashkoor, Michael Leuschel, Alexander Egyed

\end{titlepage}

\tableofcontents
\newpage

\section{Introduction}
\label{sec:introduction}
This document presents the Deliverable 1.2 of the IVOIRE project (`Formalization of VOs semantics'). 
While much work for this task was already provided by Deliverable 1.1 (`Classification of existing VOs \& tools') \cite{ivoire11}, we have not discussed \emph{refinement}, a fundamental aspect of formal methods, especially in the IVOIRE project which also affects the VOs' semantics.
This report discusses how refinement and validation relate to each other.
In general, we have two main goals:

\begin{itemize}
\item We would like to integrate formal validations (realized by VOs) in a refinement-based software development process
\item We would like to use the idea of refinement to apply validations to different domain experts' views.
\end{itemize}

To tackle the first goal, we will look at how validations and thus VOs evolve during the refinement-based software development process.
We introduce the non-linear refinement approach, which includes the abstraction of software models, to tackle the second goal. This also aims to ease the application of different validation techniques.

This report is structured as follows: \cref{sec:recap} recaps the refinement idea of IVOIRE. In~\cref{sec:refinement} we introduce the idea behind (VO) refinement and its variation. In \cref{sec:non-linear-approach}, we will introduce the non-linear refinement approach and conclude the report with a demonstration in~\cref{sec:demo}.

\section{Recap of the IVOIRE Idea}
\label{sec:recap}

\begin{figure}[t]
    \centering
    \includegraphics[width=12cm]{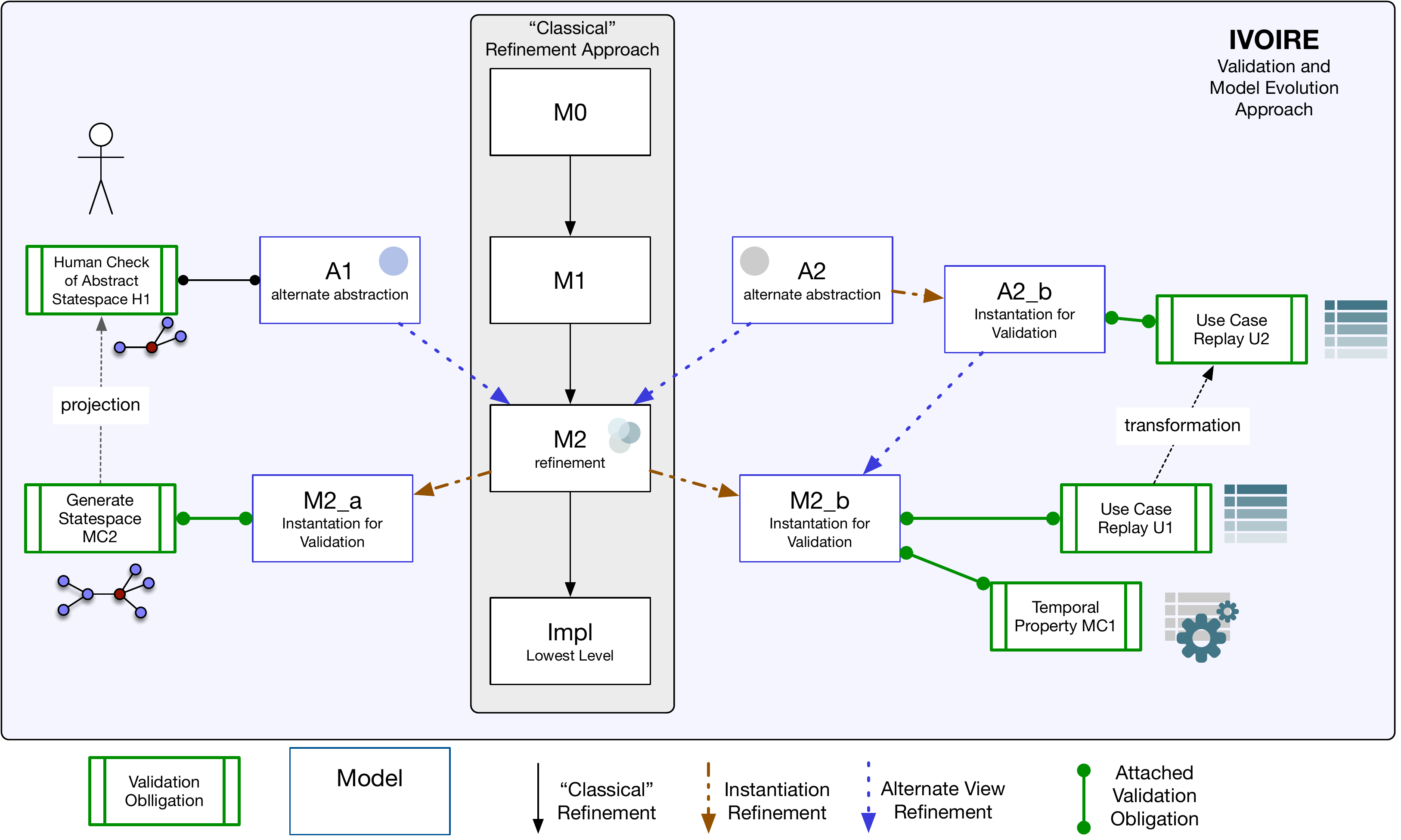}
    \caption{Refinement-based Software Development Process with VOs}
    \label{fig:software-process-vos}
\end{figure}

In Deliverable 1.1, we have proposed a systematic approach to validate requirements with \emph{validation obligations} (VOs).
We have also presented a formalization of VOs and a classification of the underlying validation techniques.
By definition, a VO validates the desired requirement and therefore shows certain behavior in the model.
\cref{fig:software-process-vos} shows the VO-based software development process in the context of IVOIRE.

Incrementally enriching the model to the implementation level is the state-of-the-art approach for formal languages such as B, Event-B, ASM, and TLA+. An abstract model is refined multiple times within the development process until it encodes all requirements. 
Additionally, it is often desired to reach an implementation level close to the final code. 
The IVOIRE approach caters to this, as illustrated by the grey box in \cref{fig:software-process-vos}.
In addition to the classical refinement chain, IVOIRE also considers non-linear refinements, which are composed of:
\begin{itemize}
    \item \textbf{Instantiation} (red dotted line).
        Instantiating a model with specific parameters has two main tasks: (1) to test and simulate the model with specific configurations the domain expert is interested in, and (2) to shrink the state-space to make model checking techniques easier to use.
        Instantiation refinement is well-known and implicitly supported by classical refinement.
    \item \textbf{Abstraction} (blue dotted line).
        We want to extend the linear refinement to allow multiple abstract views (from the perspectives of different stakeholders),
        which would free us from the strict hierarchy of the refinement chain and would open the opportunity to view a model from different aspects.
        This aligns with the notion of validation, which is the concept of checking a specification from different points of view, as pointed out by \textcite{rushby93FAA}. Abstractions need a deeper investigation which we conduct in \cref{subsec:abstractions}.
    \item \textbf{VO transformation}.
        A VO may be transformed to one another to better adapt to a changing model. There are, however, challenges to the idea that we will lay out in \cref{subsec:vo_transformation}.
\end{itemize}

Each refinement step in the model needs to ensure its consistency with previous steps.
Proving techniques (via proof obligations) can be used here, e.g. as they are applied in Rodin for Event-B.
There are also explicit-state exploration approaches, e.g. FDR, to verify the model's refinement steps.
We will consider both techniques to ensure the consistency of non-linear refinement steps.

In the following we discuss how to adapt VOs along the non-linear refinement chain as illustrated in \cref{fig:software-process-vos}.
The basic motivation is to ensure that observed behavior (what can the model do?) in M1 of \cref{fig:software-process-vos} is also present in M2.

\section{Refinement}
\label{sec:refinement}
In the context of validation, we define refinement as the enhancement or extension of functionality.
Resulting from this, refinement in the VO context allows dealing with the extension of functionality and behavior in the model while preserving the VOs ability to represent the requirement it was created for.

In the following, we will describe the relationship of a requirement to a VO and its components.
By understanding the interaction of the different parts, we can better understand how we refine the VO.
Therefore, in the following, we discuss the different directions of potential VO refinement and why we decided on the parameter-based approach, which we demonstrate later in \cref{sec:demo}.

\begin{figure}[t]
    \centering
    \begin{subfigure}[b]{0.48\textwidth}
    \includegraphics[scale=0.5]{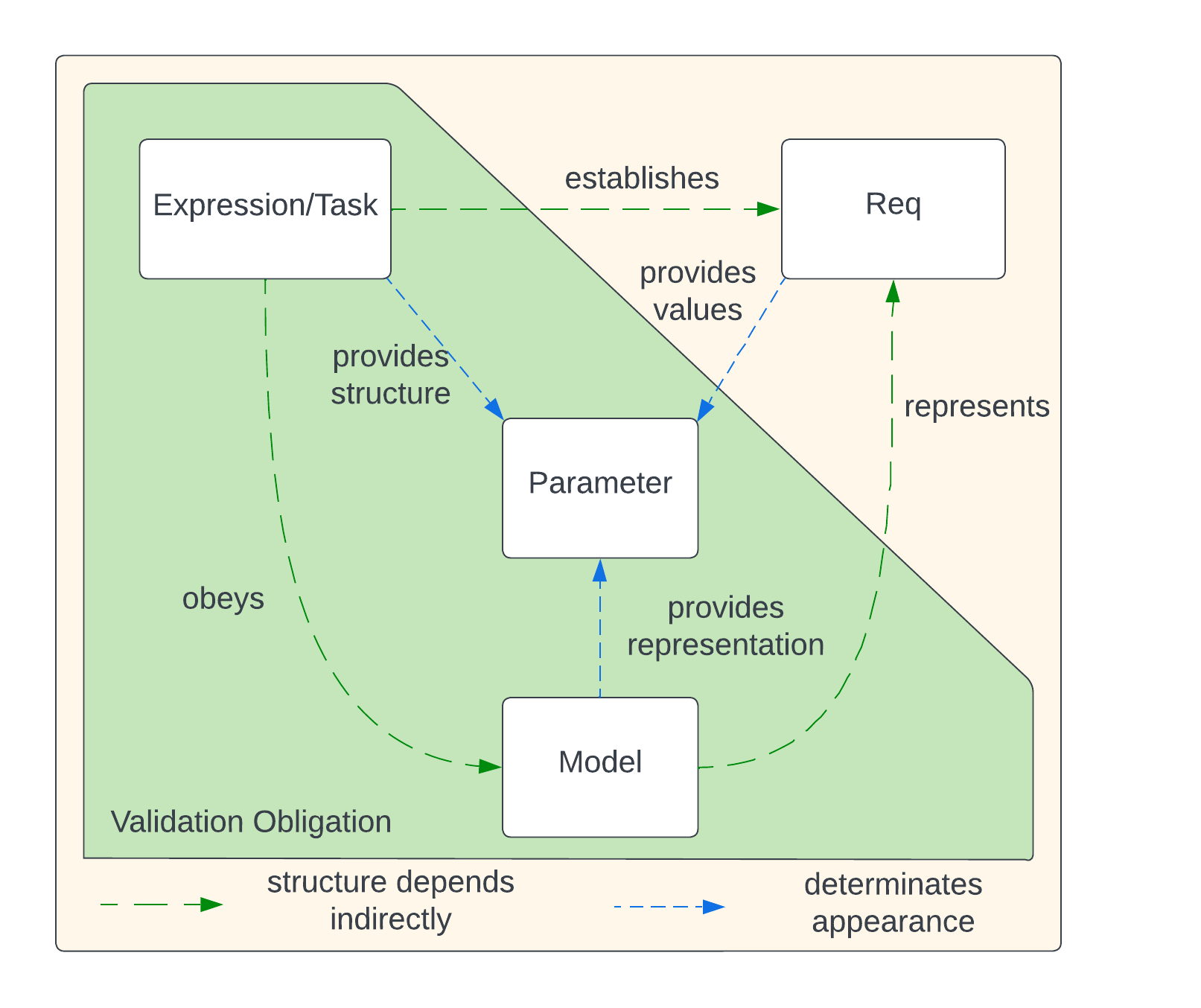}
    \caption{Dependency Relation between Requirement, Validation expression, Validation Task Parameter, and Model}
    \label{fig:dependenciesVOs}
    \end{subfigure}
    ~
    \begin{subfigure}[b]{0.48\textwidth}
    \includegraphics[scale=0.35]{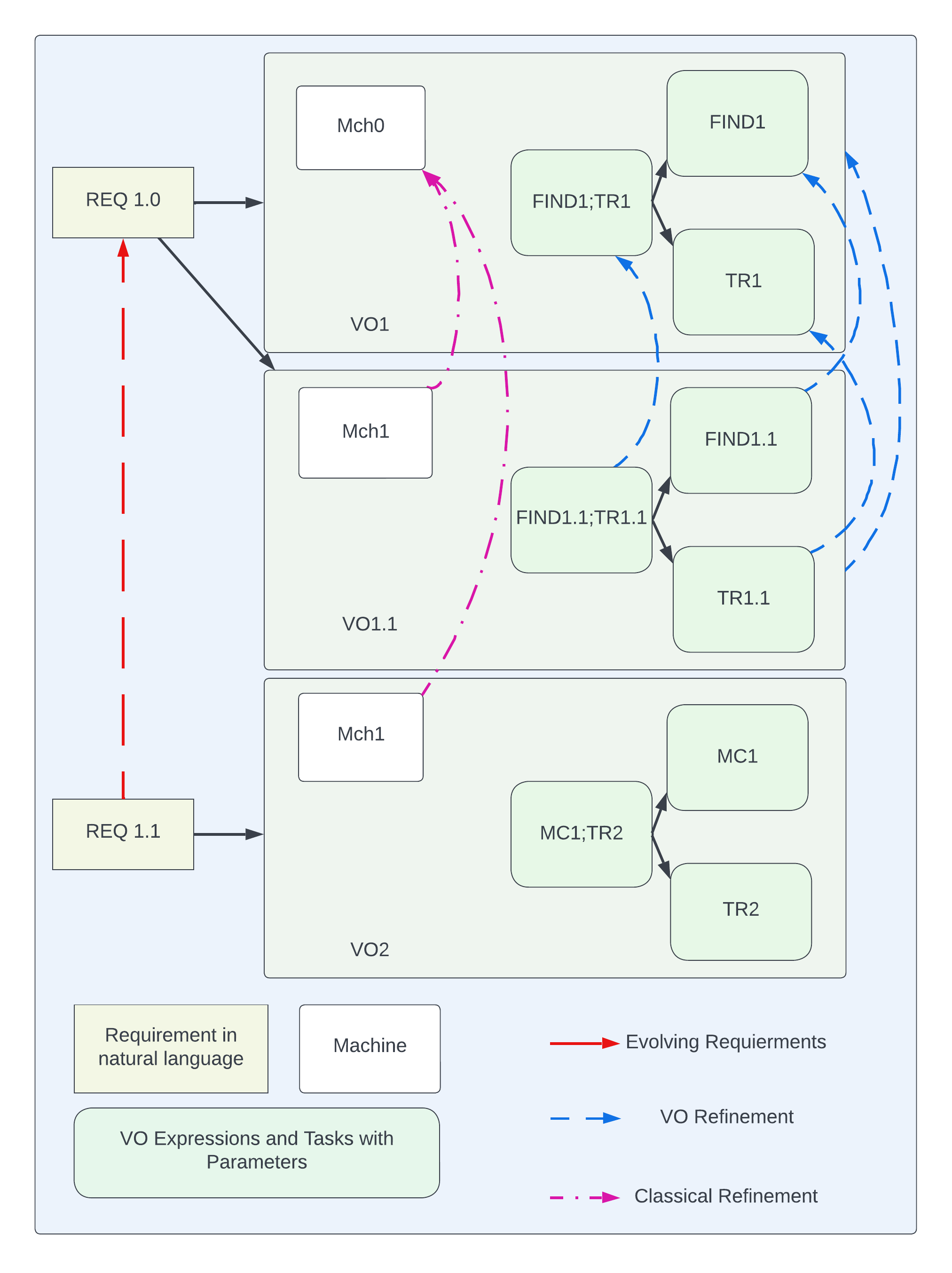}
    \caption{Relationship between VO Refinement, Requirement Refinement, and Classical Refinement}
    \label{fig:voRefinement}
    \end{subfigure}
    \caption{Dependencies between internal VO components and VO refinement chain}
\end{figure}

In \cref{fig:dependenciesVOs}, we can see that a requirement is directly responsible for a validation expression's composition, i.e., the requirement has a strong influence on the expression, which represents a formalization of the requirement and also on the actual task that has to be executed. 
The model is encoded to satisfy the requirement of a formal representation, and the validation expression is indirectly dependent on how the requirement is encoded into the model.
The indirect coupling gives the modeler the freedom to choose the appropriate task for the problem. 
However, we will see later in this section that this freedom also has a disadvantage.

Requirements, model, and validation expression are responsible for shaping the appearance of the parameters for the validation tasks involved:

\begin{enumerate}
    \item The requirement determines which values need to be checked.
    \item The model determines the appearance of the values under investigation.
    \item The task determines the appearance of the parameters.
\end{enumerate}

As we now look into the relationship of VO components, let us observe how formal models are evolved. For this we consider \cref{fig:voRefinement}. We have three major components here: requirements (yellow), models (white), and VOs (green). In the following, we discuss the evolution of each of these components in the context of formal refinement.

The pink arrow symbolizes traditional refinement from one machine to another. The blue arrow symbolizes the refinement of a VO, and the red arrow symbolizes the refinement of a requirement.

What might be surprising is that in this representation, the refinement of VOs is decoupled from the refinement of requirements. The rationale behind this becomes clear when we take another look at \cref{fig:dependenciesVOs}. Two cases can happen in ongoing development. First, (1) the model can evolve without changing the requirement. Second, (2) the requirements evolve, thus affecting the model and, therefore, the task with which the requirement is represented. In the following we will explain (1) in more detail in \cref{subsec:vo_refinement} and (2) in \cref{subsec:reqRef}.

\subsection{VO Refinement}
\label{subsec:vo_refinement}
We start with the simpler (1) case, which is represented by the blue arrows in \cref{fig:voRefinement}. The refinement effect for a model is already known and understood and poses no scientific challenge.

If the model changes but not the requirement, this change affects the model's contribution to the validation task parameters. However, as we understand refinement for models, we can apply this understanding to the parameter needed by the VO and contributed by the model. In short, we can apply similar refinement rules to the parameters of the VO as we do in model refinement.

For example, variables subject to data refinement in Event-B are coupled via a gluing invariant. Therefore, we can use this gluing invariant to refine a task's parameters. In addition, whenever a parameter uses a variable that is about to be replaced in a model refinement step, we can use the gluing invariant to apply the replacement in the parameter.
Another example would be LTL formulas. Several modeling languages allow extending the set of allowed behavior in a refinement step by allowing new, concrete transitions. An LTL formula would need to be extended to include those new transitions while not compromising on the expressiveness of the abstract behavior. For Event-B, \textcite{hoang2016foundations} give an algorithm and examples for LTL refinement.
Also, variables used in the LTL formula need to be adapted using the gluing invariant.
The change of a model is strictly formalized and follows defined rules.
We can apply these rules to the parameters that are themselves formalized, similar to the model.

In conclusion, if the model is refined, we need to keep the VO intact by applying the changes the model has undergone to the representation the model provides.
The task itself remains unchanged.
This might sound very simple, but actually poses quite a few challenges.
For example, we know that we need to translate transitions and variables, but currently, we have no way of automatically doing it. 
It is currently up to the modeler to find the right translation, which is error-prone and can be challenging. One challenge in particular is translating predicates into expressions. Gluing invaraints are written as predicates while VO parameters often need expressions. 
Therefore, in the future we need to come up with automatic translation. One starting point could be the work of~\textcite{konrad2011translation}.
We will see an example of a manual translation in \cref{subsec:abstraction_refinment_chain}.

\subsection{Evolving Requirements}
\label{subsec:reqRef}
However, in case (2), represented by the red arrow in \cref{fig:voRefinement}, we have a different situation. Regarding (1), we ensure that the requirement in its current form is preserved and re-established in a changing environment, but one might want to enhance or modify the requirement along with the model. 
We have two options: (a) to `refine' the validation expression/task to mirror the enhanced refinement, or (b) to create an enhanced version of the requirement and deliberately create a new VO for this requirement.

\subsubsection{Refining the Validation Expression}
\label{subsubsec:refining_ve}
The problem with (a) is that we do not have a refinement calculus for validation tasks. 
As explained in Deliverable 1.1~\cite{ivoire11}, we decided to rely on the modeler to find the appropriate task for a requirement. 
This follows the rationale that validation itself can never be fully proven but is a matter of finding perspectives on the model that shows what we want to validate, as pointed out by \textcite{rushby93FAA}:

\begin{quote}
    By their very nature, the problems of validating top-level specifications or statements of assumptions do not lend themselves to definitive proof; all that can be done is to examine the specifications from various perspectives, and to probe and test their consequences until we are satisfied that they adequately capture our requirements, or the real-world phenomenon they are intended to describe.
\end{quote}

Previously, we decided that it was the modeler's responsibility to choose an appropriate point of view (task) to show the requirement's presence. This was based on the assumption made in an earlier draft of Deliverable 1.1~\cite{ivoire11} that a total formalization of requirements might be possible but very difficult. However, as inspired by Rushby, we have corrected our assumption. 
Formalizing requirements that need to be validated is not possible in the well-defined manner that applies to verification.

If we now change the requirements, it will affect the representation in the form of VOs. 
However, we run into a major problem due to the missing well-defined formalization. 
We cannot know for sure if, how, and in which way the VO and its underlying components will change. 
With this insight gained, we dismiss approach (a).

\subsubsection{Creating New Validation Expressions}
In (b), the modeler creates an enhanced version of the requirement alongside the existing one. From this one, the modeler can create a new VO that represents the requirement. The newly created requirement contains the elements of the old one and, in addition, the newly added ones.
Resulting from this, we have two VOs for the same model: VO1.1 for the original requirement and VO2 for the enhanced requirement, as can also be seen in~\cref{fig:voRefinement}.
Assuming that both VOs are implemented correctly, the relationship between both VOs is $VO1.1 \subset VO2$, meaning that VO2 shows everything that VO1.1 does, plus additional features. Therefore, we are using $\subset$ in the mathematical sense to describe the relationship as one including the other.
When we create VO2 to cater to an enhanced requirement, we might encounter a situation where this VO2 has different tasks than VO1.1. For example, in \cref{fig:voRefinement} we replaced the task of finding a certain state with complete model checking of the state space.
A consequence of evolving requirements can be that VOs get 'abandoned' because they are not further refined as the things they represent are now part of another VO. E.g., in \cref{fig:voRefinement} we have created a VO2 for which $VO1.1 \subset VO2$ is true. Therefore to ensure REQ1 and REQ1.1, we just carried on with VO2 and did not further cater for VO1.1.

Therefore, our decision is to only allow addition to requirement and their enhancement. Additionally, this fits the spirit of formal methods and refinement. We only add to the specification but never remove it from it. The same is true for our view of requirements. If we want to subtract things from the requirements we have introduced into our model, why should we have introduced such things first?

\subsection{VO Transformation}
\label{subsec:vo_transformation}
With the notion of $VO1.1 \subset VO2$ and is associated meaning of inclusion, the idea of $VO1.1 \subseteq VO2$ and therefore the possible equality $VO1.1 = VO2$ of VOs is close. Introducing equality between VOs means that we can transform one VO into another, fully preserving the intention encoded by the VOs. The difficulty is that we have to take care that both VOs express the same thing, as pointed out for the $VO1.1 \subset VO2$ case in~\Cref{subsec:reqRef}. A VO transformation is beneficial when we want to switch out a task as we might become aware that a task is better suited to fulfill our needs than the one we currently use, for example, when we replace a trace with an LTL formula. However, as pointed out earlier, in~\cref{subsubsec:refining_ve} we have no well-defined formalization meaning that translation between expressions is undefined. Therefore for finding a transformation, we would rely on the modeler's ability to find the appropriate translation while we cannot assist with a formalism. 

The effect of a VO transformation for $VO1.1 = VO2$  would be that we can decide which VO we want to keep in further refinement steps.

\section{Non-Linear Refinement Approaches}
\label{sec:non-linear-approach}
As discussed in \cref{sec:recap}, non-linear refinement is particularly important for the IVOIRE approach. 
The goal is to open up the refinement chain from a strict top-down approach to a more flexible approach.
In addition to the classical refinement chain, we consider \emph{alternate view abstraction} and \emph{instantiation view refinement}.
Based on a detailed model, \emph{alternate view abstraction} intends to abstract the model focusing on components that are interesting for a domain expert.
\emph{Instantiation view refinement} aims to provide a specific instantiation for testing and simulation purposes for a domain expert.

Some specification languages like CSP offers different levels of how one process can refine another for checking trace, failure, failure-divergence refinement properties as pointed out by~\textcite{derrick18a}. Such refinement relations can be also proofed automatically with FDR\cite{fdr}. What is however particular useful in CSP is the ability to have multiple 'parents' for one process which is particular useful in two cases:
\begin{itemize}
    \item One can merge multiple concepts that were observed individually before.
    \item One is not forced to use an artificially made up topology for the features, which can be a problem as it can be challenging for the interaction between individual components. 
\end{itemize}
As a result, allowing multiple parents can increase flexibility concerning validation.
In contrast, languages like B and Event-B are more constrained in their refinement process. The refinement relationship gets established, usually by proofs (in Event-B) or checking if all traces of the concrete specification are possible in the abstract one (B). Multiple parents would increase the complexity of this endeavor drastically.

Nevertheless, observing individual parts and their combination in a model might be desired, and there is an approach to open up this strict hierarchy called decomposition, which was proposed by \textcite{abrial2007refinement}.
The original idea was to split one specification into multiple other ones to easier reason about the sub-components.
The approach also provides a technique to put the split components back together later in development.
However, the decomposition approach, in general, has a disadvantage when it comes to the treatment of variables shared between sub-components.
To de-compose machines, all variables shared between the sub-components become static, meaning that refinements of the sub-components cannot replace shared variables (data refinement) or manipulate their invariants.
This is crucial to re-composition the sub-components without conflicts.
One simply imagine the case where the original machine has an invariant \texttt{@inv1 $a : 1..5$} is de-composition into two sub-components \texttt{A} and \texttt{B}. 
If we would refine \texttt{A} and introduce a new invariant \texttt{@inv1.A $a : 3..5$} and we refine \texttt{B} and introduce \texttt{@inv1.B $a : 1..3$}, we would be not able to re-compose both sub-specifications, because the new invariants would conflict.

From this example, we can follow that the weak point of decomposition is aligning conflicting invariants when merging sub-components back together. This is also true for the multiple parents in a refinement. In our refinement calculus, we need to define how to deal with conflicts while ensuring correct refinement.
To refine multiple parents into one child, both parents must be fully independent regarding their variables. In this case, the combination would be trivial. 
In all other cases, we would need to find and resolve potential conflicts if this is even possible.

In \cref{subsec:abstractions} we will introduce the concept of \emph{alternate view abstraction}, also called \emph{abstract views} or simply \emph{abstractions}, to Event-B. The idea is to offer the main advantages a multiple parent approach offers, i.e., observing components and their combinations independent from the topological sorting while evading the problems of conflicting invariants and behaviors.
Regarding VOs, in our reasoning, the semantics introduced so far regarding creation and refinement stay intact.

In \cref{subsec:init_vo} we will introduce the second non-linear refinement approach, namely \emph{instantiation view refinement}. This one is not new; however, we will define how VOs behave in this case. In an \emph{instantiation view refinement}, also called \emph{instantiation view} or \emph{scenario view}, we create an instance of the model and reason about this instance. This can have advantages for understanding the model, as instances remove abstractness and foster understanding.

\subsection{Alternate View Abstraction}
\label{subsec:abstractions}
While a classical refinement step encodes more details, abstraction reduces the details in the model by focusing on some components only.
The main purpose is to abstract the model for a stakeholder, and to apply higher-level validations that can be traced back to the concrete model. By creating an abstraction we can rid ourselves from the artificially imposed topology and can observe components isolated from the rest of the model.

In \cref{subsubsec:projection_approach} we will first discuss the approaches we explored to create an abstraction, namely projection and abstract interpretation. 

In \cref{subsubsec:inverting_classical_refinement} we present our solution for creating abstractions and discuss how abstractions affect refinement of VOs.

\subsubsection{Projection Approaches}
\label{subsubsec:projection_approach}
Initially, we assumed that the main reason for abstracting a model is that it contains too many variables and events that are not relevant to the stakeholder.
We concluded that by simply removing the accessed variables, we create an abstraction that is useful for modeling purposes. For this, we had two main approaches:

\begin{itemize}
\item \textbf{Model projection} was the first approach we considered. It is based on \emph{projection}, i.e., hiding/removing irrelevant variables and events while leaving others in place.
Hiding, in this case, means that every check on the state space would be performed as usual, but the modeler/user only sees the variable they specified.
So, for example, a state only consisting of hidden variables would no longer be visible or interactable.
In the same way, if a transition reaches a state that is identical to the current state modulo hidden variables, that transition would be hidden.

The idea was to focus on specific variables and events
using an approach similar to state-space projection
as presented by \textcite{DBLP:conf/icfem/LadenbergerL15}.
However, this is only possible if the state space is finite and relatively small, which resulted in us dismissing this approach.

\item \textbf{Abstract interpretation} \cite{cousot} as an approach was derived from the idea that we want to reason about an abstract version of the model.
Abstract interpretation would allow more efficient dealing with larger or possibly infinite state spaces.
However, there are two major disadvantages when deploying abstract interpretation in the VO and modeling context. 
\begin{itemize}
    \item First, and most important is that abstract interpretation is also a mere projection and does not allow changes to the fundamental structure of the model.
    \item Creating abstract interpretation requires additional skills, i.e., finding and creating an projection expression. Such an expression might become complicated.
    \item We would need to translate the VO parameters for VOs shown on the abstraction back. This would happen via the projection expression. However, as the projection is not an equivalence this translation might introduce additional ambiguity.
\end{itemize}
In the end we dismissed abstract interpretation mainly because of the reason that it does not offer the ability to change the model.
\end{itemize}

A common disadvantage of both approaches is that we need to implement a state-space representation. In the case of projection, we would need to implement a state-space representation on which we then perform validation/verification operations like LTL checking. With the abstract interpretation approach, we first need an algorithm to produce the abstraction. Second, we also need a state-space representation of this abstraction on which we can run queries.
In both cases, the problem is that state-space representation should align with the current one so that we can use existing infrastructure to reason about properties.
In conclusion, we found that projection-based approaches suffer from three problems:

\begin{enumerate}
\item They are not flexible enough when dealing with complex models with many variables.
\item They do not allow to decide freely about the content we want to put into the abstraction.
\item They require the modeler to learn a new technique.
\item We are currently missing infrastructure to produce and reason about the state-space resulting from such an abstraction. Providing such infrastructure is sizeable effort.
\end{enumerate}

The previous section focused on taking an abstract point of view when working with a model. However, what changed our approach entirely was the insight that an abstract view might not be enough. It might be necessary to replace existing concrete variables with simpler abstracted variables and simplify the model's events accordingly to better understand how components interact with each other.

\subsubsection{Inverting Classical Refinement}
\label{subsubsec:inverting_classical_refinement}
As projection has severe disadvantages, we propose a different approach which we summarize as \textit{inverting classical refinement}.
The advantages are that we can create views onto the model the way we tried with projections. Additionally, we can introduce new simple abstract variables that represent more complicated concrete ones.
We found this incredibly useful when reasoning about the model and giving a demonstration (see \cref{sec:demo}). 
The second advantage is that we do not need to worry about finite state spaces or the unintended loss of information. 
The third advantage is that we do not use new keywords and thus do not need to implement a new mechanic from scratch. This reduces implementation effort compared to the other proposed approaches. We might decide to implement abstractions explicit in the future to ease the work with this technique. However, such an implementation would be a mere visual representation, and the correct mechanism would remain untouched. Therefore only limited effort is needed.
We can conclude that everything we need for our approach already exists.
A drawback is that much repetitive work has to be done by hand, but we are optimistic that we can automate these parts in the future.

\begin{figure}
    \centering
    \begin{subfigure}[b]{0.48\textwidth}
        \centering
        \includegraphics[scale=0.35]{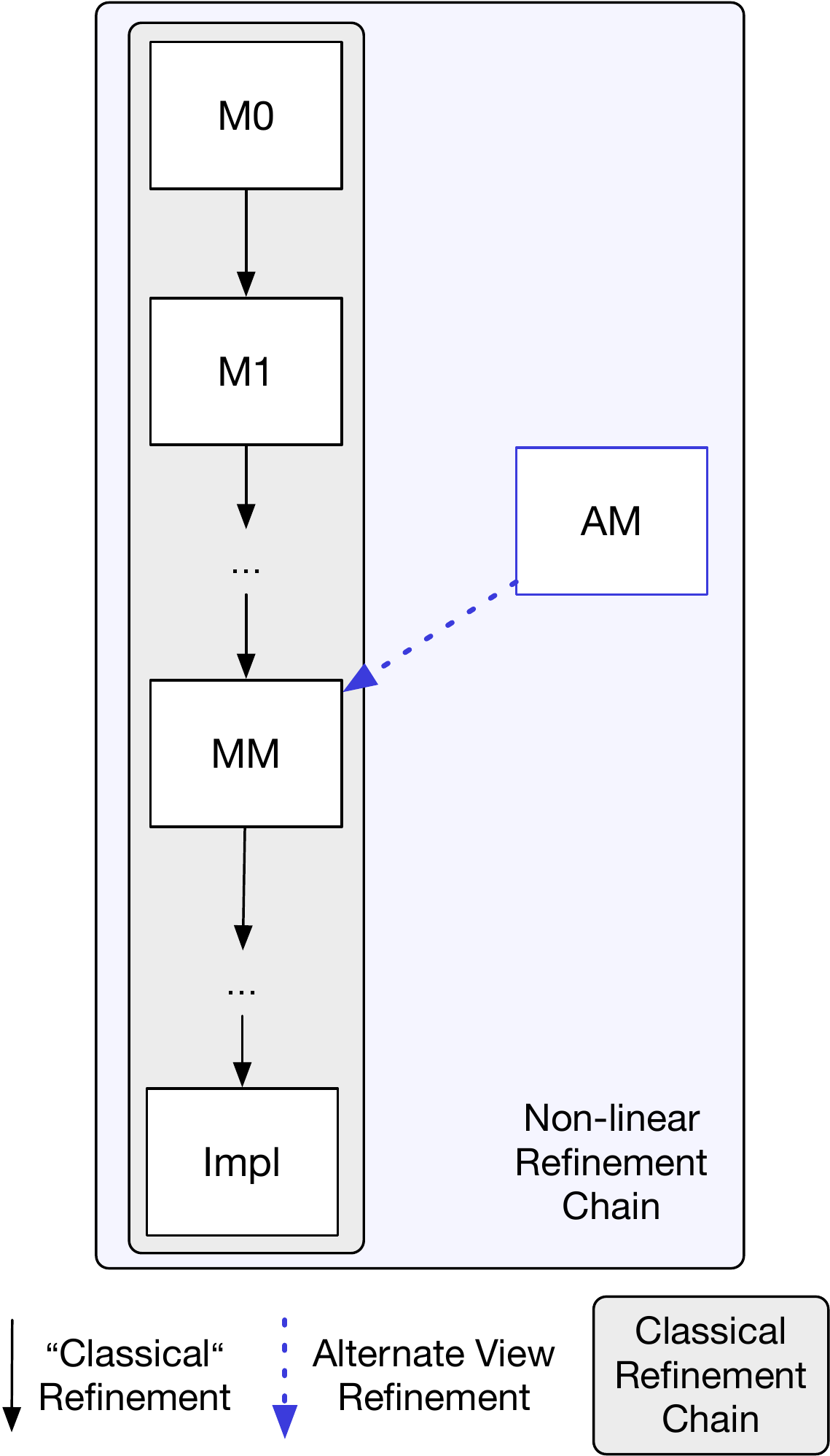}
        \caption{Refinement Chain with Abstraction}
        \label{fig:abstraction}
    \end{subfigure}
    ~
    \begin{subfigure}[b]{0.48\textwidth}
        \centering
        \includegraphics[scale=0.35]{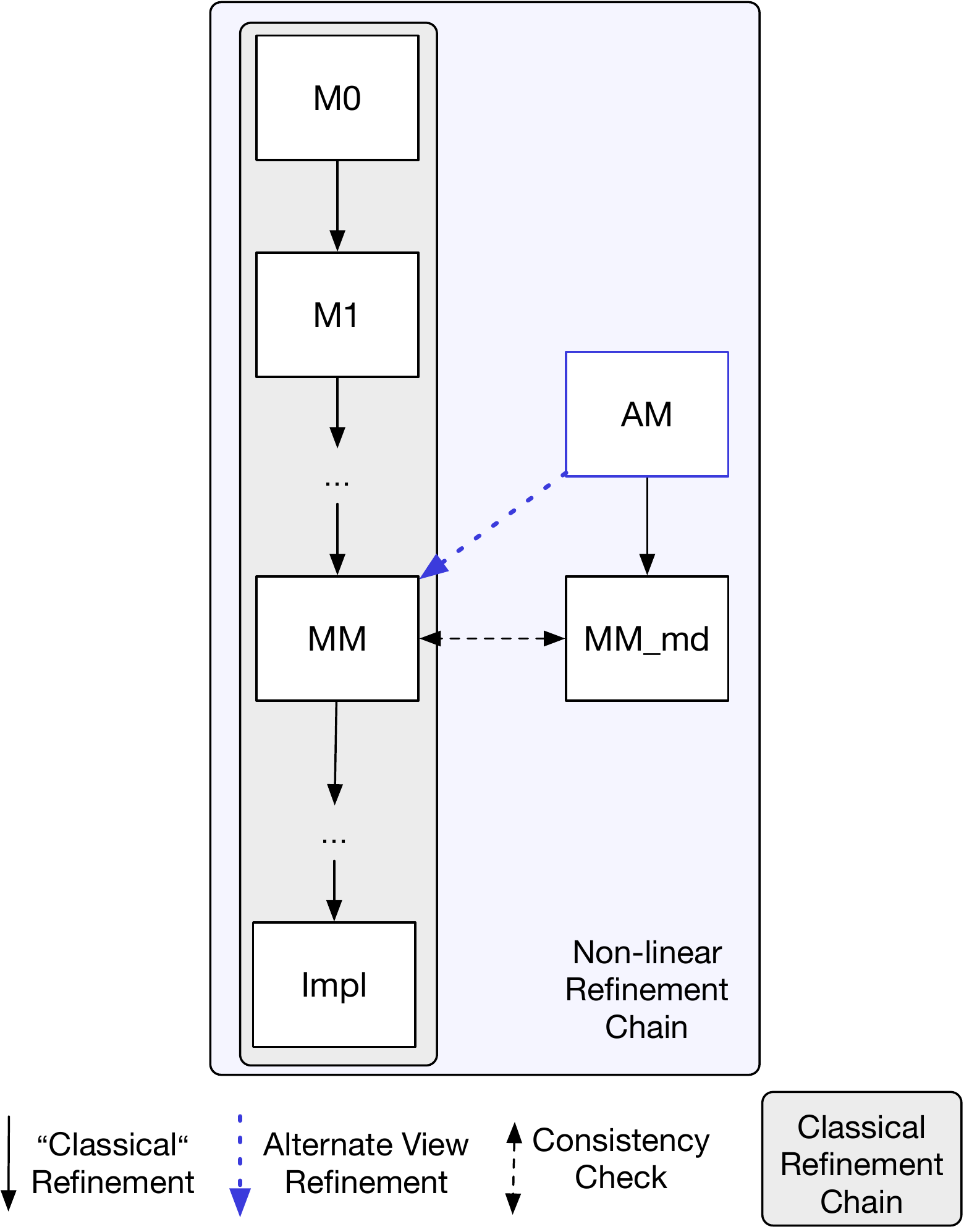}
        \caption{Abstraction Internally}
        \label{fig:abstraction-internal}
    \end{subfigure}
\end{figure}

Let us consider a refinement chain with an abstraction step as portrayed in~\cref{fig:abstraction}.
On the left-hand side, there is a classical refinement chain from \texttt{M0} to \texttt{Impl}.
Our interest is now to derive an abstraction \texttt{AM} from \texttt{MM}, which is located in the middle of the refinement chain.
To our best knowledge, the concept of alternate view abstraction is novel for formal methods.
However, the novelty raises the problem that the semantics of an alternate view abstraction is not defined yet.

In the context of linear refinement, the term `abstraction' is already used by, e.g., Atelier~B~\cite{clearsy_b_refman} to refer to the machine that is being refined by a refinement machine.
That is, \texttt{M1} is an abstraction of \texttt{M2} iff \texttt{M2} is a refinement of \texttt{M1}.
We extend this term to non-linear refinement: if \texttt{AM} is an alternate view abstraction of \texttt{M}, then \texttt{M} must be a refinement of \texttt{AM}.
This definition makes it possible to use existing refinement techniques, including refinement consistency checks.

Thus, to prove that \texttt{AM} is a valid abstraction of \texttt{M1}, we must show that \texttt{M1} can be derived from \texttt{AM} via refinement.
As mentioned in the introduction for \cref{sec:non-linear-approach}, this causes problems in many refinement-based formalisms, including the Event-B method, where a model can only refine a single parent. Our abstraction would violate this rule, as we need to prove the correct refinement relationship for two parents without them contradicting each other.

\paragraph{Creation of Abstractions}
In \cref{fig:abstraction-internal}, we propose a solution for this problem. 
First, we flatten the refinement chain leading from \texttt{M0} down to \texttt{MM} and create a flattened machine \texttt{MM\_md}. 
Flatten means that we rewrite the indirect representation of the refining machine into a single machine representing all machines leading down to \texttt{MM}.
With the refinement chain flattened, \texttt{MM\_md} can then refine the alternate abstraction \texttt{AM} and establish the correctness of the refinement.

The next step is to prove the refinement relationship between \texttt{AM} and \texttt{MM\_md}. 
As with classical refinement, the modeler may need to define and add gluing invariants to link the abstract variables in \texttt{AM} with their concrete counterparts in \texttt{MM\_md}.

In the final step, it must be shown that \texttt{MM\_md} behaves identically to the original \texttt{MM}. 
This is crucial for the VOs from \texttt{AM} being transferable to \texttt{MM}.

We need to check one thing to ensure identical behavior: Every new insertion has to be a gluing invariant. The gluing invariant has to follow the information available in the flattened machine. We are not allowed to introduce new variables or axioms as this would enrich \texttt{MM\_md} compared to \texttt{MM}. If we have only gluing invariants created from existing information, we can carry over our assumption to the original machine as we assured that no additional information was added. With this, we have established that indeed \texttt{AM} would be a correct parent of \texttt{MM}.

\paragraph{Technical Aspects}
Technically we create an abstraction by letting the modeler create \texttt{AM}.
Then, the user builds \texttt{MM\_md} and shows the refinement relationship with \texttt{AM}.
Currently, the absence of additional additions has to be checked manually.

In the future, we would like to implement first-class support for abstractions in the B method so that a modeler can declare an \emph{abstraction} relationship between two machines.
This would require adding a new keyword, e.g. \texttt{abstracts}, and the ability to specify gluing invariants inside the abstraction machine instead of the refinement.
With proper tool support, e.g., proof obligation generation for Rodin, we would like to avoid generating temporary machines to prove the abstraction relationship.

\subsubsection{Effect on VOs}
\label{subsubsec:effectInstantiationonVOs}
The rules to prove that an abstraction is a valid abstraction for an existing model is analogous to proving that the existing model is a valid refinement for the abstraction. $\square$
Following from this, we can apply the fact that abstraction is analogous to refinement to VOs. Therefore, the rules of VO refinement apply to those VOs we created on an abstraction and want to show on the original model.

\subsection{Instantiation Refinement}
\label{subsec:init_vo}
The concept of instantiation is not novel and has already been used for many years. Instantiation focuses on replacing abstract values with concrete values, thus allowing to reason about this specific scenario. Instantiation does not extend the feature set of a model.
For example, a railway interlocking model might be designed generically, independent of any particular track layout.
One could then instantiate the model with a concrete real-life track topology (for details, see \cref{subsec:instant_ref}).
The advantage is that we can show this to stakeholders who are familiar with this example from real life, and we can explain how our model works on this well-known and understood example.

\subsubsection{Connection with Abstractions}
In \cref{fig:software-process-vos}, one can see that \texttt{A2\_b} is an instance of an abstraction. 
In this case, all previously defined rules apply. We can treat the instance of an abstraction similar to an instance created from a regular model.
Consequently, for \texttt{M2\_b}, \texttt{A2\_b} is an abstraction and one can use the gluing invariant used for establishing the abstraction relationship to show the VOs from \texttt{A2\_b} on \texttt{M2\_b}. 

\subsubsection{Effect on VOs}
\label{subsubsec:effectAbstractionVOs}

The previously formulated rules for the refinement of VOs also apply in the context of instances.

\begin{figure}
    \centering
    \includegraphics[scale=0.5]{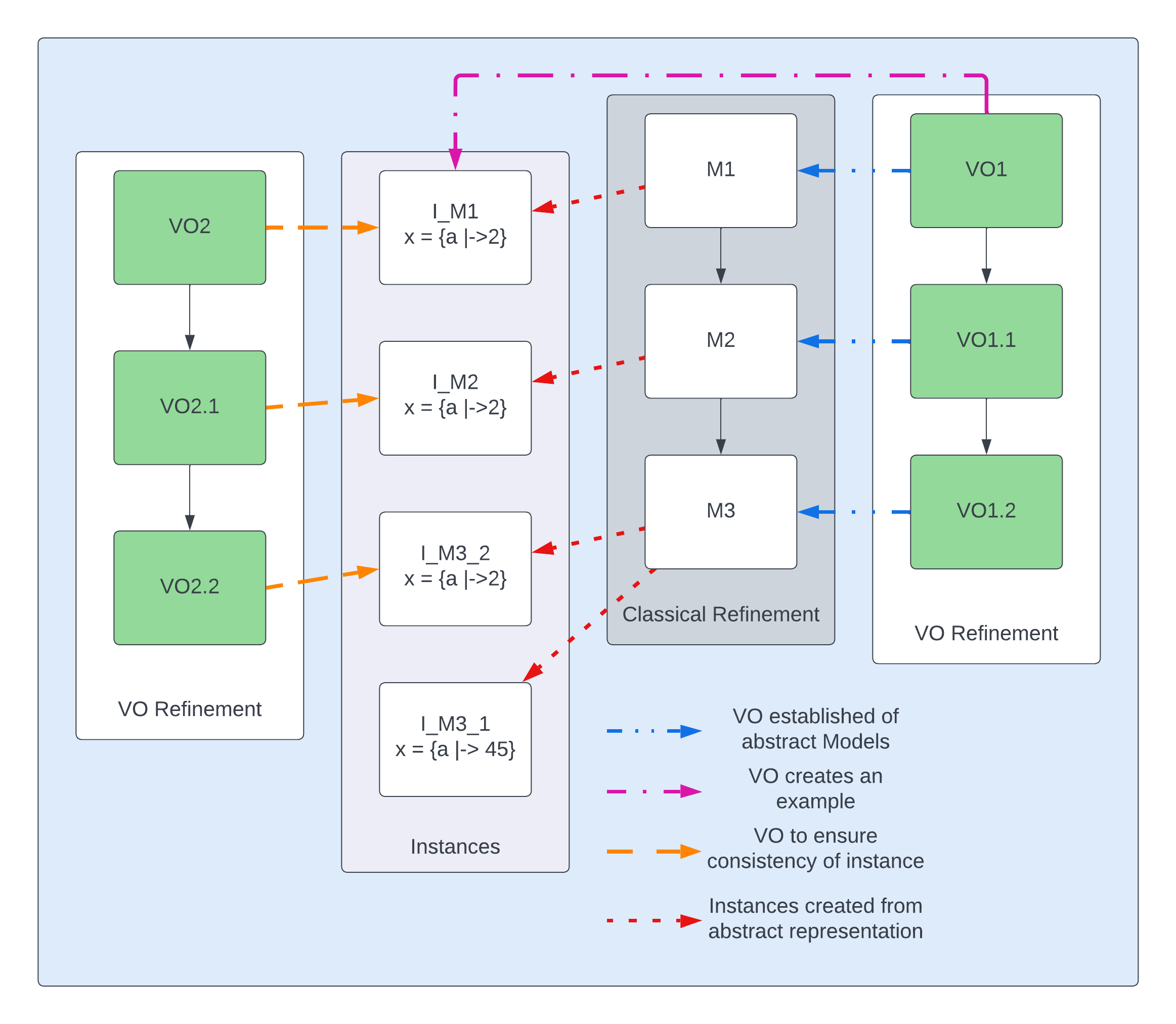}
    \caption{Instances and their relationship to VOs}
    \label{fig:instances_and_vos}
\end{figure}

Instances offer several application opportunities with different goals. Thus goals dictate how the VO is treated. Consider \cref{fig:instances_and_vos}. In the center we have a refinement chain \texttt{M1} to \texttt{M3}. We have a \texttt{VO1} that we established on \texttt{M1} and show this VO successively for \texttt{M2} and \texttt{M3} (blue arrows). In the context of instantiation, we can use VOs in two different ways. 
First, a VO for an abstract model can be applied to an instance and, this way, create scenarios (pink arrow). 
This is dependent on the task, however.
Especially animation is a suitable task for creating examples while proving is not necessarily a good task as proving (PO task) works mostly with abstract properties.
Second, VOs might be created to ensure that a given scenario on an instance always works (orange arrow). 
Naturally, whenever we create the scenario by replacing an abstract value with a concrete one, we might want to ensure that this scenario holds everywhere, which we ensure with a VO. 
This has an interesting effect when refining the model. Imagine we established \texttt{VO2} on \texttt{I\_M1}. However, we want to keep this scenario around as it benefits our stakeholders.
Whenever we refine our model, e.g., create \texttt{M2} and use the same concrete values we used in \texttt{M1} to create the instance, then our VO needs to be refined alongside to ensure that our scenario is still valid. 

Observing \cref{fig:instances_and_vos} yields an additional insight: We could draw refinement arrows between the instances \texttt{I\_M1} to \texttt{I\_M3}!
In theory, those are proper refinements of each other as they use the same concrete values.
However, the use of this would be limited.
We would need to properly establish this refinement relationship alongside the existing one from \texttt{M1} to \texttt{M3} that we would need to establish in the process of modeling.

\subsection{Abstractions and Existing Approaches}
\label{sec:relatedWork}
Abstraction is not as powerful as having the ability to have multiple parents for one refinement. 
This is because, in the case of multiple parents, those parents are independent, and the child is their refinement and therefore has to fold to the restrictions the parents present themselves with.
Abstractions are just a different view of the existing problem created because the goal is already present, but we need to find a different abstract version.
Abstraction can therefore be seen as a second refinement with the same goal. The outcome of the refinement remains unchanged. 
However, the starting point is different and offers us insights into the model.

\section{Demonstration of Non-Linear Refinement and VO Refinement}
\label{sec:demo}

In this section, we will demonstrate (1) how to refine a VO and (2) how non-linear refinement works together with VOs.
In particular, we will show the purpose of abstraction and instantiation refinement and how to refine VOs.
For demonstration purposes, we will use the interlocking model by \textcite{abrial2010modeling}\footnote{The full code of the example can be accessed in~\cite{vu22b}}.

\begin{figure}[t]
    \centering
    \includegraphics[scale=0.38]{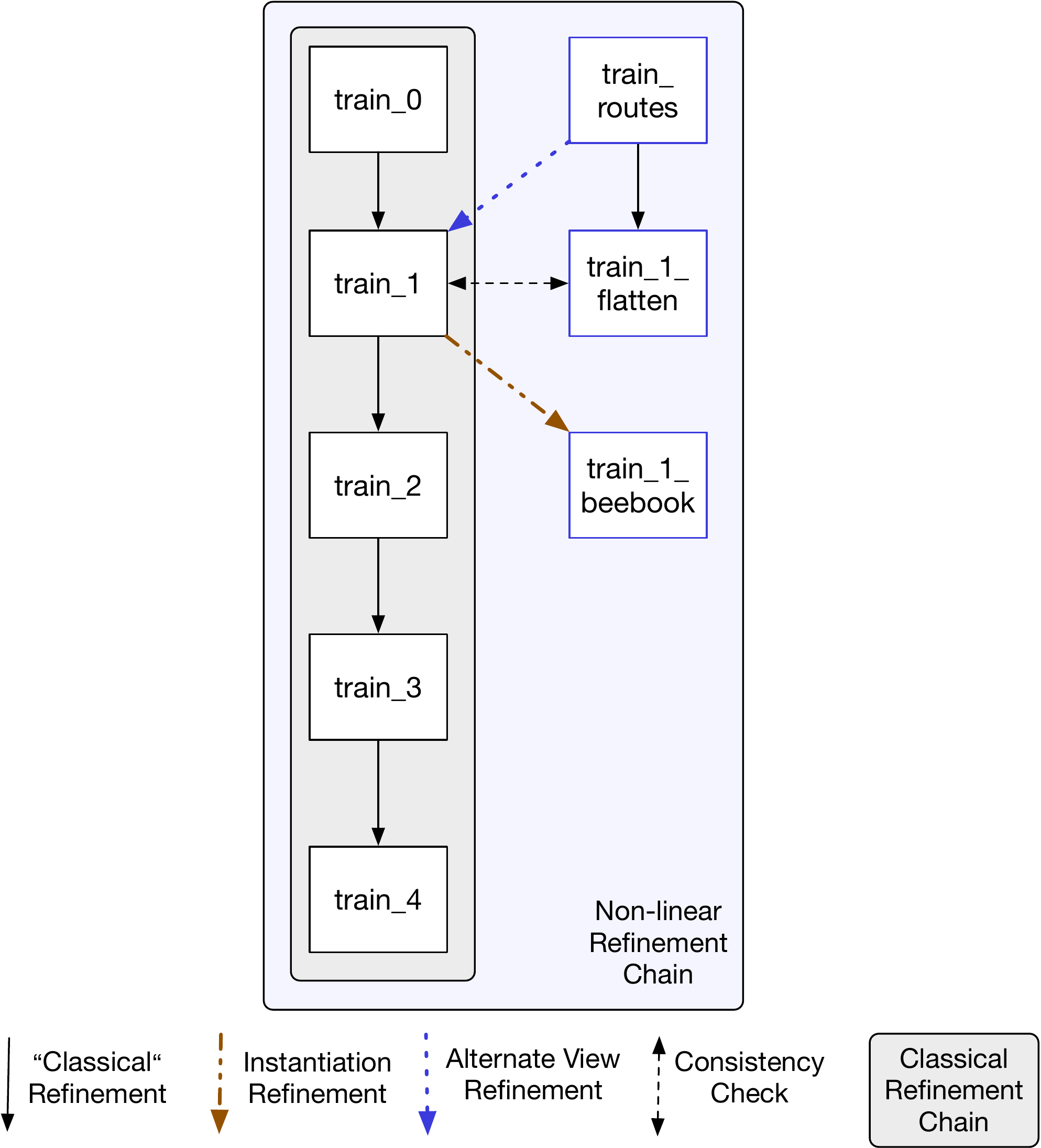}
    \caption{Non-linear Refinement Chain for Train Model}
    \label{fig:nonlinear-refinement-train}
\end{figure}

\Cref{fig:nonlinear-refinement-train} shows the refinement chain for our model.
The classical refinement chain is as follows:
\begin{itemize}
    \item \texttt{train\_0} is the abstract model which introduces routes and blocks. Within the abstract model, routes can be reserved and freed. It also tracks which blocks are occupied by a train.
    \item \texttt{train\_1} introduces point positioning and route formation as intermediate steps between the reservation of a route and the corresponding train's movement.
    \item \texttt{train\_2} introduces a variable specifying ready routes, i.e., routes that are ready to accept a train. 
    \item \texttt{train\_3} refines this variable by introducing signals.
    \item In the specifications \texttt{train\_0} through \texttt{train\_3}, the interlocking topology is kept generic. \texttt{train\_4} finally instantiates the interlocking topology with concrete blocks and routes.
\end{itemize}

Regarding the classical refinement chain, we will focus on \texttt{train\_0} and \texttt{train\_1}.

In our VO demonstration, we will also introduce new requirements that Abrial does not consider. 
Abrial formulates his requirements to demonstrate verification by proof obligations.
Thus, we also try to formulate requirements related to validation than verification.
In addition to classic VO refinement, we introduce an abstraction \texttt{train\_routes} and an instantiation \texttt{train\_1\_beebook}.
For the VOs, we will show how to refine a VO from \texttt{train\_0} to \texttt{train\_1}, we will create a VO on \texttt{train\_routes} and show a property not easily observable on the original model, and we show how to refine the VO back to \texttt{train\_1}. 
Finally, we show how to use VOs in the context of instantiations with \texttt{train\_1\_beebook}.

To sum up, we are focusing on:

\begin{itemize}
\item \texttt{train\_0} and \texttt{train\_1} in the classical refinement chain
\item \texttt{train\_routes} as an abstraction of \texttt{train\_1}
\item \texttt{train\_1\_beebook} as an instantiation of \texttt{train\_1}
\end{itemize}

This is also illustrated in \cref{fig:nonlinear-refinement-train}.

\subsection{VO Refinement}
\label{subsec:vorefinementExample}
To demonstrate VO refinement, we consider the following requirement given by a (non-technical) stakeholder:

\begin{itemize}
    \item REQ1: A route can be reserved. A train can enter and pass this route, and afterwards the route can be freed.
\end{itemize}

This is a very high-level statement for a multitude of reasons. 
First, the requirements request that all potential routes fulfill the property. 
Second, in \texttt{train\_0}, we do not have actual routes but reason over the abstract idea of routes. 
Third, and this hides behind the ambiguity of words: We want to know if this behavior is possible in any way. We have no contraints like it should be possible \texttt{G}lobally; meaning infinite times often. We simply want to know if it is possible at all.
Such a requirement is often encountered when a person tries to understand the model. 
This is also an example of why animation is useful, as it allows us to explore the model manually. 
Finally, as \texttt{train\_0} is very generic, we want to gain confidence in the model, to check that the requirement REQ1 is fulfilled.
For this, we create an arbitrary trace with the help of the ProB animator, and check if REQ1 holds.
For this, we create VO1 which is applied to \texttt{train\_0}.
VO1 consists of a single validation task TR1:
\begin{itemize}
    \item VO1: TR1
    \item TR1/\texttt{train\_0}/TR: \texttt{trace}
\end{itemize}

Now, we want to check if this general requirement about the model is true for the model's refinement.
Therefore, we refine the trace for \texttt{train\_1}.
The refined VO and VT are called VO1.1 and VT1.1, respectively.
According to our findings in \cref{subsec:vo_refinement}, we refine the validation task's parameter \texttt{trace}.
This is done via trace refinement, which results in a refined trace and, therefore, a refined parameter.

\begin{figure}
    \centering
    \includegraphics[scale=0.2]{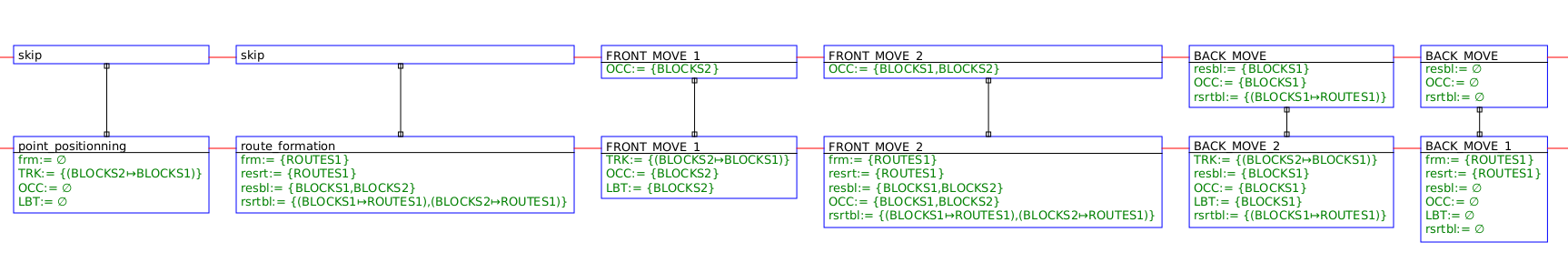}
    \caption{Original and refined trace of the VO}
    \label{fig:traceRef}
\end{figure}

The result of the trace refinement can be seen in \cref{fig:traceRef}; for space reasons, we omitted the unchanged parts.
The upper part is a piece of the original trace, and the lower part is from the refined trace. \texttt{skip} indicates where transitions had to be inserted to adapt the trace.
Namely \texttt{point\_positioning} and \texttt{route\_formation} had to be newly inserted. 
This is acceptable and does not violate our requirement as this is concrete, unrelated behavior.
Another change is that \texttt{BACK\_MOVE} was split into two events \texttt{BACK\_MOVE\_1} and \texttt{BACK\_MOVE\_2} as the train representation has become more detailed.
This is also not a problem as it does not violate our requirements.
As the refinement was successful, we conclude that the abstract behavior remains intact.
Furthermore, we are satisfied with the newly added transitions in the expected bounds.

\subsection{Instantiation Refinement Chain}
\label{subsec:instant_ref}
\begin{figure}[ht]
    \centering
    \includegraphics[scale=0.45]{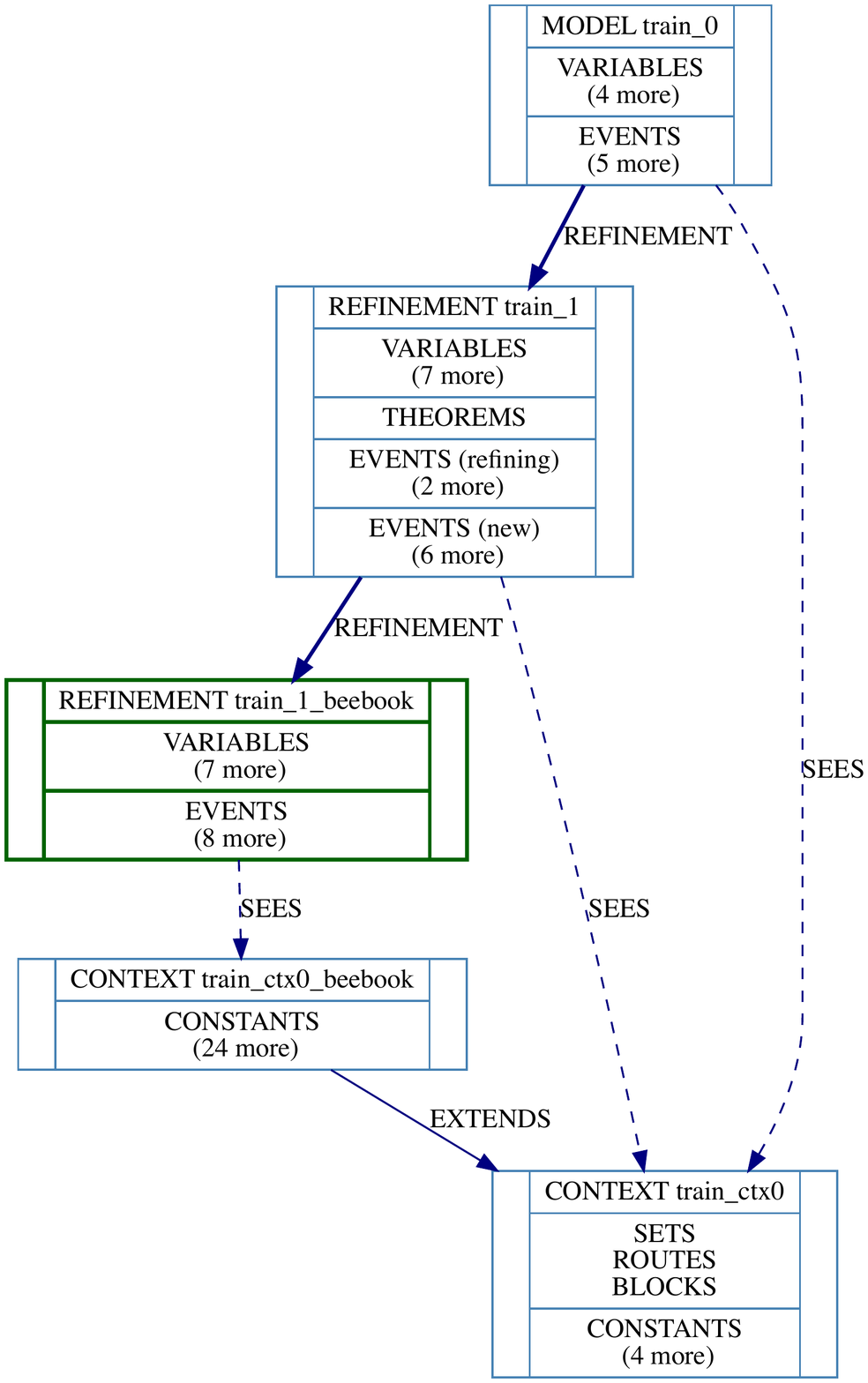}
    \caption{Train Instantiation Machine Hierarchy}
    \label{fig:train-instantiation-machine-hierarchy}
\end{figure}

Based on \texttt{train\_1}, we would like to run scenarios on a concrete instantiation of the interlocking topology.
Therefore, we will instantiate \texttt{train\_1} with the interlocking topology as described in~\cite{abrial2005b}
which results in the component \texttt{train\_1\_beebook}.
\cref{fig:train-instantiation-machine-hierarchy} shows the instantiation refinement chain from \newline \texttt{train\_0} to \texttt{train\_1\_beebook},
consisting of the corresponding machines and contexts.
Here, one can see that \texttt{train\_1\_beebook} is a refinement of \texttt{train\_1}, with both machines \emph{seeing} different contexts.

\begin{figure}[ht]
    \centering
    \includegraphics[scale=0.45]{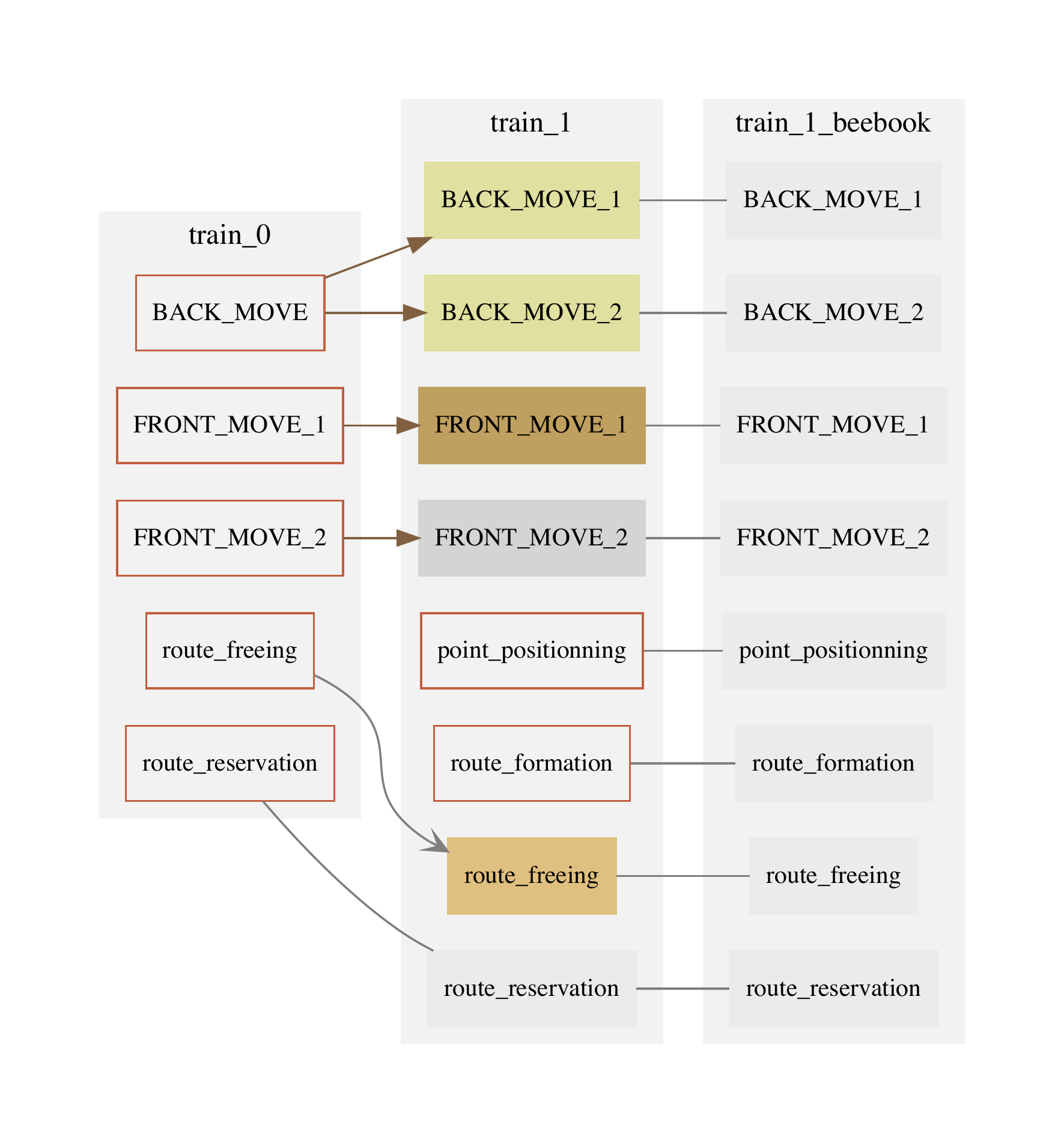}
    \caption{Train Instantiation Event Hierarchy}
    \label{fig:train-instantiation-event-hierarchy}
\end{figure}

\cref{fig:train-instantiation-event-hierarchy} shows the event refinement hierarchy of the instantiation refinement chain.
Events in \texttt{train\_1} are refined one-by-one by those in \texttt{train\_1\_beebook}.
\Cref{lst:b-train-1-beebook} even shows that those events are only extended, without adding any features.
The main difference is the inclusion of the different contexts.
As shown in \cref{fig:train-instantiation-event-hierarchy}, \texttt{train\_1}
and \texttt{train\_1\_beebook} \emph{sees} \texttt{train\_ctx0} and \texttt{train\_ctx0\_beebook} respectively.
Based on both contexts, we will demonstrate how the instantiation works.

\begin{figure}[ht]
    \centering
    \includegraphics[scale=0.45]{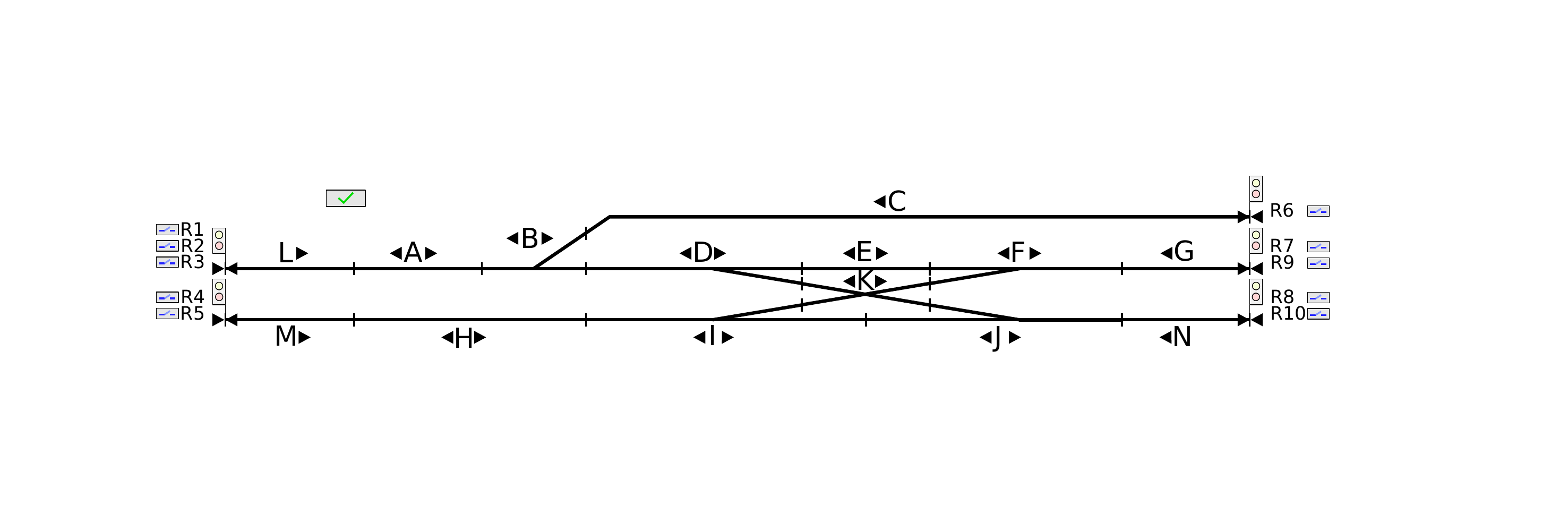}
    \caption{Domain-specific View of a Train Instantiation}
    \label{fig:train-instantiation-domain-specific}
\end{figure}

\Cref{lst:b-train-ctx-0} shows the context \texttt{train\_ctx0} which defines axioms for a generic interlocking topology.
The blocks and routes are primarily defined as deferred sets, i.e., they are specified generically.
Furthermore, the connection between the blocks and their corresponding routes are also specified generically.
\cref{lst:b-train-ctx-instantiation} defines an instantiation for the routes and blocks.
Here, one can see the definition of 10 routes (\texttt{R1} to \texttt{R10}), 14 blocks (\texttt{A} to \texttt{J}), and how the blocks are connected to the respective routes.
A domain-specific view for the instantiated interlocking topology is shown in \cref{fig:train-instantiation-domain-specific}.

\newpage

\begin{lstlisting}[
caption={train\_1\_beebook},escapechar=§,label={lst:b-train-1-beebook}]
machine train_1_beebook refines train_1 sees train_ctx0_beebook 
variables resrt resbl rsrtbl OCC TRK frm LBT 
events
  event INITIALISATION extends INITIALISATION end
  event route_reservation extends route_reservation end
  event route_freeing extends route_freeing  end
  event FRONT_MOVE_1 extends FRONT_MOVE_1 end
  event FRONT_MOVE_2 extends FRONT_MOVE_2  end
  event BACK_MOVE_1 extends BACK_MOVE_1  end
  event BACK_MOVE_2 extends BACK_MOVE_2  end
  event point_positionning extends point_positionning end
  event route_formation extends route_formation end
end
\end{lstlisting}

\newpage

\begin{lstlisting}[
caption=train\_ctx0,
escapechar=§,
label=lst:b-train-ctx-0]
context train_ctx0
sets ROUTES BLOCKS
constants rtbl nxt fst lst
axioms
  @axm1 rtbl §$\in$§ BLOCKS §$\rel$§ ROUTES
  @axm2 dom(rtbl) = BLOCKS
  @axm3 ran(rtbl) = ROUTES
  @axm4 nxt §$\in$§ ROUTES §$\tfun$§ (BLOCKS §$\pinj$§ BLOCKS)
  @axm5 fst §$\in$§ ROUTES §$\tfun$§ BLOCKS
  @axm6 lst §$\in$§ ROUTES §$\tfun$§ BLOCKS
  @axm7 fst~ §$\subseteq$§ rtbl
  @axm8 lst~ §$\subseteq$§ rtbl
  @axm11 §$\forall$§r.r §$\in$§ ROUTES §$\implies$§ fst(r) §$\neq$§ lst(r) 
  @axm10 §$\forall$§r.r §$\in$§ ROUTES §$\implies$§ (§$\forall$§S.S §$\subseteq$§ ran(nxt(r)) §$\wedge$§ S §$\subseteq$§ nxt(r)[S] §$\implies$§ S = §$\emptyset$§)
  @axm9 §$\forall$§r.r §$\in$§ ROUTES §$\implies$§ 
                   nxt(r) §$\in$§ rtbl~[{r}] §$\setminus$§ {lst(r)} §$\pinj$§ rtbl~[{r}] §$\setminus$§ {fst(r)} 
  @axm12 §$\forall$§r,s.r §$\in$§ ROUTES §$\wedge$§ s §$\in$§ ROUTES §$\wedge$§ r §$\neq$§ s §$\implies$§ 
                  fst(r) §$\notin$§ rtbl~[{s}] §$\setminus$§ {fst(s),lst(s)} 
  @axm13 §$\forall$§r,s.r §$\in$§ ROUTES §$\wedge$§ s §$\in$§ ROUTES §$\wedge$§ r §$\neq$§ s §$\implies$§ 
                  lst(r) §$\notin$§ rtbl~[{s}] §$\setminus$§ {fst(s),lst(s)} 
end
\end{lstlisting}

\begin{lstlisting}[
caption=train\_ctx0\_beebook,
escapechar=§,
label=lst:b-train-ctx-instantiation]
context train_ctx0_beebook extends train_ctx0
constants A B C D E F G H I J K L M N
axioms
  @axm44 partition(BLOCKS, {A}, {B}, {C}, {D}, {E}, {F}, {G}, {H}, {I},{J}, 
          {K},{L},{M},{N})
  @compute_rtbl_from_nxt rtbl = {b §$\mapsto$§ r §$\mid$§ r §$\in$§ dom(nxt) §$\wedge$§ 
         (b §$\in$§ dom(nxt(r)) §$\vee$§ b §$\in$§ ran(nxt(r)))}
  @axm40 nxt = {(R1 §$\mapsto$§ {L §$\mapsto$§ A, A §$\mapsto$§ B,B §$\mapsto$§ C}),
         (R2 §$\mapsto$§ {L §$\mapsto$§ A, A §$\mapsto$§ B,B §$\mapsto$§ D,D §$\mapsto$§ E,E §$\mapsto$§ F, F §$\mapsto$§ G}),
         (R3 §$\mapsto$§ {L §$\mapsto$§ A, A §$\mapsto$§ B,B §$\mapsto$§ D,D §$\mapsto$§ K,K §$\mapsto$§ J, J §$\mapsto$§ N}),
         (R4 §$\mapsto$§ {M §$\mapsto$§ H,H §$\mapsto$§ I,I §$\mapsto$§ K,K §$\mapsto$§ F,F §$\mapsto$§ G}),
         (R5 §$\mapsto$§ {M §$\mapsto$§ H,H §$\mapsto$§ I,I §$\mapsto$§ J,J §$\mapsto$§ N}),
         (R6 §$\mapsto$§ {C §$\mapsto$§ B,B §$\mapsto$§ A,A §$\mapsto$§ L}),
         (R7 §$\mapsto$§ {G §$\mapsto$§ F,F §$\mapsto$§ E,E §$\mapsto$§ D,D §$\mapsto$§ B,B §$\mapsto$§ A,A §$\mapsto$§ L}),
         (R8 §$\mapsto$§ {N §$\mapsto$§ J,J §$\mapsto$§ K,K §$\mapsto$§ D,D §$\mapsto$§ B,B §$\mapsto$§ A,A §$\mapsto$§ L}),
         (R9 §$\mapsto$§ {G §$\mapsto$§ F,F §$\mapsto$§ K,K §$\mapsto$§ I,I §$\mapsto$§ H,H §$\mapsto$§ M}),
         (R10 §$\mapsto$§ {N §$\mapsto$§ J,J §$\mapsto$§ I,I §$\mapsto$§ H,H §$\mapsto$§ M})}
  @axm41 fst = {(R1 §$\mapsto$§ L),(R2 §$\mapsto$§ L),(R3 §$\mapsto$§ L),
         (R4 §$\mapsto$§ M),(R5 §$\mapsto$§ M),
         (R6 §$\mapsto$§ C),
         (R7 §$\mapsto$§ G),(R8 §$\mapsto$§ N),
         (R9 §$\mapsto$§ G),(R10 §$\mapsto$§ N)}
  @axm42 lst = {(R1 §$\mapsto$§ C),(R2 §$\mapsto$§ G),(R3 §$\mapsto$§ N),
         (R4 §$\mapsto$§ G),(R5 §$\mapsto$§ N),(R6 §$\mapsto$§ L),
         (R7 §$\mapsto$§ L),(R8 §$\mapsto$§ L),
         (R9 §$\mapsto$§ M),(R10 §$\mapsto$§ M)}
end
\end{lstlisting}

\paragraph{Example}

In our model, routes can be used simultaneously without restriction if they are not conflicting. Let us imagine a situation where a domain expert does not trust the model. He proposes an example rail topology (\texttt{train\_1\_beebook}) for which he knows all non-conflicting routes, and he wants us to show that our implementation can achieve the expected results. 

\begin{itemize}
    \item REQ2: the following routes do not conflict with each other: route 1 and route 10, route 2 and route 5, ...
\end{itemize}

This is a very high-level requirement that consists of multiple sub-requirements. 
Usually, to satisfy this requirement, one has to satisfy all sub-requirements. 
A VO validating this requirement would apply a set of regression tests, testing all possibilities. 
However, for the sake of example, we do not test all routes. Instead, we concentrate on one pair to show instantiation abilities. 
The two routes are fully independent of each other, meaning they do not share blocks. 
If our model is correct, we can move two trains on each route; there should not be any conflict between them. 
We propose a VO in the form of a trace that ensures consistency for all involved pairs of routes. 
This VO can be satisfied if we generate a trace for each constellation.

\begin{itemize}
    %\item VO2:TR(trace1) \& TR(trace2) \& ..
    \item VO2: TR2.1 \& TR2.2 \& \ldots
    \item TR2.1/\texttt{train\_1\_beebook}/TR: \texttt{trace1}
    \item TR2.2/\texttt{train\_1\_beebook}/TR: \texttt{trace2}
\end{itemize}

\begin{figure}[htbp]
    \centering
    \includegraphics[scale=0.45]{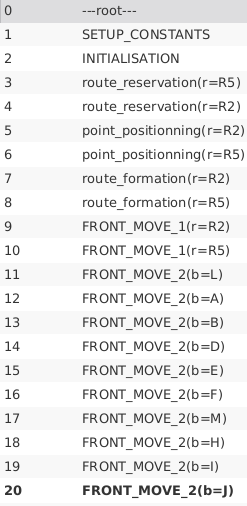}
    \caption{Trace in VO2}
    \label{fig:trace_VO}
\end{figure}

\begin{figure}[t]
    \centering
    \includegraphics[scale=0.45]{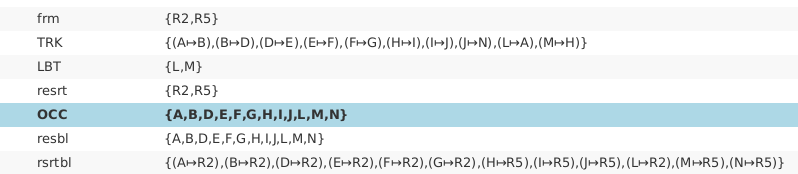}
    \caption{Variable constellation in the last step of the trace}
    \label{fig:variables}
\end{figure}

\Cref{fig:trace_VO} shows a trace to validate the sub-requirement showing the conflict-freeness of route 2 and route 5. 
The whole length of each route is occupied by the train, as we can see in~\cref{fig:variables}. 
In order to check this behavior, we apply a postcondition check in the trace replay task.

\newpage

\subsection{Abstraction Refinement Chain}
\label{subsec:abstraction_refinment_chain}
Based on this refinement level where blocks and routes are introduced,
we want to validate each route's behavior on a more abstract level.
Here, the desired sequence for each route is: route reservation, route formation, route freeing.
To achieve this, we will abstract \texttt{train\_1} to \texttt{train\_routes} as shown in \cref{lst:b-train-3}.
Within the abstracted model, there is only one variable which describes for each route whether it is \emph{free}, \emph{formed}, or \emph{reserved}.
The abstracted status of a route is specified by a new set \texttt{RoutesStatus} which is defined in \texttt{train\_routes\_ctx} (see \cref{lst:b-train-ctx-abstract}).

\newpage

\begin{lstlisting}[
caption=train\_routes,
escapechar=§,
label=lst:b-train-3]
machine train_routes sees train_routes_ctx
variables rs
invariants @inv_routes rs §$\in$§ ROUTES §$\tfun$§ RoutesStatus
events
  event INITIALISATION then 
    @init_Routes rs := ROUTES §$\times$§ {free} 
  end
  
  event route_reservation
    any r where
      @grd_Routes_1 r §$\in$§ ROUTES
      @grd_Routes_2 rs(r) = free
    then
      @act_Routes rs(r) := reserved
  end
  event route_freeing
    any r where
      @grd_Routes_1 r §$\in$§ ROUTES
      @grd_Routes_2 rs(r) = formed
    then
      @act_Routes rs(r) := free
  end
  event route_formation
    any r where
      @grd_Routes_1 r §$\in$§ ROUTES
      @grd_Routes_2 rs(r) = reserved
    then
      @act_Routes rs(r) := formed
  end
end
\end{lstlisting}

\begin{lstlisting}[
caption=train\_routes\_ctx,
escapechar=§,
float,
label=lst:b-train-ctx-abstract]
context train_routes_ctx
sets RoutesStatus ROUTES
constants free reserved formed
axioms 
  @axm_r partition(RoutesStatus, {free}, {reserved}, {formed})
end
\end{lstlisting}

As Event-B only allows refinement of a single machine,
we will introduce a machine \texttt{train\_1\_flatten} where \texttt{train\_0} and \texttt{train\_1} are merged.
The resulting machine is shown in \cref{lst:b-train-1-merged}.
In \cref{lst:b-train-1-merged}, we omit events, variables, and invariants that are not relevant for the routes' behavior.
As mentioned before, \texttt{train\_1\_flatten} is specified as a refinement of \texttt{train\_routes}.

Comparing both machines, one can see that the variables, invariants, and events in \texttt{train\_routes} are less complex than in \texttt{train\_1\_flatten}.
To check that \texttt{train\_routes} is an abstraction of \texttt{train\_1\_flatten}, it is necessary to check that \texttt{train\_1\_flatten} is a correct refinement of \texttt{train\_routes}.
Therefore, we have introduced the \emph{gluing invariants} \texttt{glue1}, \texttt{glue2}, and \texttt{glue3} in \texttt{train\_1\_flatten}.
Furthermore, the consistency between \texttt{train\_1} and \newline \texttt{train\_1\_flatten} must be checked, i.e., we check that the behaviors of both machines correspond to each other.

Concerning \texttt{train\_ctx0}, we will also have the adapt the context slightly (shown in \cref{lst:b-train-ctx-0-changed}).
It is now specified as an extension of the abstracted context \texttt{train\_routes\_ctx}.
As \texttt{ROUTES} is now defined in \texttt{train\_routes\_ctx}, it is removed from \texttt{train\_ctx0}.
Compared to the original version of \texttt{train\_ctx0}, the constants and axioms do not change.

\begin{lstlisting}[
caption=train\_ctx0 (modified in abstraction chain),
escapechar=§,
float,
label=lst:b-train-ctx-0-changed]
context train_ctx0 extends train_routes_ctx
sets BLOCKS
constants ...
axioms ...
end
\end{lstlisting}

\paragraph{Example}
With our abstraction, we want to show that the sequence for each route's behavior is always: $\texttt{reserved} \longrightarrow \texttt{formed} \longrightarrow \texttt{freed}$.
However, as pointed out earlier, the route status is implicitly contained in the model.
Reserved and formed routes are explicitly specified, while free routes are implicitly defined as routes that are neither reserved nor formed.
To demonstrate each route's behavior more actively, we have created \cref{lst:b-train-3}. First let us define our requirement.

\begin{lstlisting}[
caption=train\_1\_flatten (merged for abstraction chain),
escapechar=§,
float,
label=lst:b-train-1-merged]
machine train_1_merged refines train_routes sees train_ctx0 
variables resrt resbl rsrtbl OCC TRK frm LBT
invariants
  @inv1.1 resrt §$\subseteq$§ ROUTES
  @inv1.2 resbl §$\subseteq$§ BLOCKS
  @inv1.3 rsrtbl §$\in$§ resbl §$\tfun$§ resrt
  @inv1.5 rsrtbl §$\subseteq$§ rtbl
  @inv2.1 TRK §$\in$§ BLOCKS §$\pinj$§ BLOCKS
  @inv2.2 frm §$\subseteq$§ resrt
  @glue1 rs~[{free}] §$\cap$§ (resrt §$\cup$§ frm) = §$\emptyset$§
  @glue2 rs~[{reserved}] = resrt §$\setminus$§ frm
  @glue3 rs~[{formed}] = frm
  ...
events
  event INITIALISATION
    then
      @act1 resrt := §$\emptyset$§
      @act2 resbl := §$\emptyset$§
      @act3 rsrtbl := §$\emptyset$§
      @act5 TRK := §$\emptyset$§
      @act6 frm := §$\emptyset$§
      ...
  end
  
  event route_reservation refines route_reservation
  any r where
    @grd1.1 r §$\notin$§ resrt
    @grd1.2 rtbl~[{r}] §$\cap$§ resbl = §$\emptyset$§
  then
    @act1.1 resrt := resrt §$\cup$§ {r}
    @act1.2 rsrtbl := rsrtbl §$\cup$§ (rtbl §$\ranres$§ {r})
    @act1.3 resbl := resbl §$\cup$§ rtbl~[{r}]
  end
  
  event route_freeing refines route_freeing
  any r where
    @grd1.1 r §$\in$§ resrt §$\setminus$§ ran(rsrtbl)
  then
    @act1.1 resrt := resrt §$\setminus$§ {r}
    @act2.1 frm := frm §$\setminus$§ {r} 
  end

  event route_formation refines route_formation
  any r where 
    @grd1 r §$\in$§ resrt §$\setminus$§ frm
    @grd2 rsrtbl~[{r}] §$\ranres$§ nxt(r) = rsrtbl~[{r}] §$\domres$§ TRK
  then
    @act1 frm := frm §$\cup$§ {r}
  end
  
  event FRONT_MOVE_1 ... end
  event FRONT_MOVE_2 ... end
  event BACK_MOVE_1 ... end
  event BACK_MOVE_2 ... end
  event point_positionning ... end
end

\end{lstlisting}

\begin{itemize}
    \item REQ3: For every route the order of the route status is always \texttt{reserved} $\longrightarrow$\texttt{formed} $\longrightarrow$\texttt{freed}.
\end{itemize}
For this we create a VO on the abstraction: \texttt{train\_routes}.

\begin{itemize}
    %\item VO3: MC;SPRJ(rs(R1)), \ldots, MC;SPRJ(rs(R8))
    \item VO3: MC(FIN);(SPRJ1 \& \ldots \& SPRJ8)
    \item MC/\texttt{train\_routes}/MC
    \item SPRJ1/\texttt{train\_routes}/SPRJ: rs(R1)
    \item \ldots
    \item SPRJ8/\texttt{train\_routes}/SPRJ: rs(R10)
\end{itemize}

\begin{figure}
    \centering
    \includegraphics[scale=0.45]{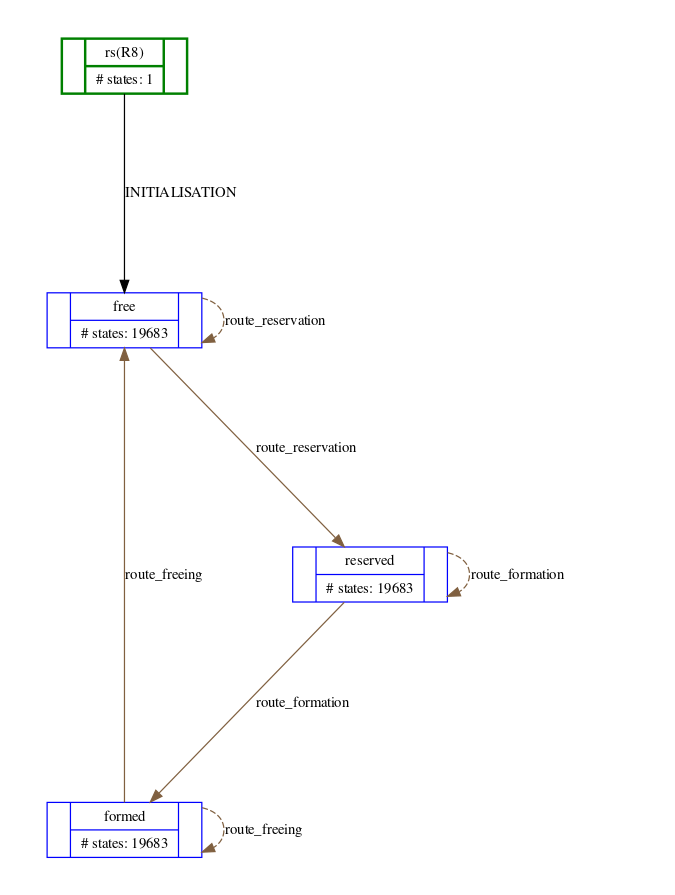}% Huch
    \caption{Projection of \texttt{train\_routes}' State Space on Route 8}
    \label{fig:abstractPrj}
\end{figure}

We uncover the state-space for the model via model checking and then create the projection onto each route to check if the route satisfies our condition.
Such a projection can be seen in \cref{fig:abstractPrj}.
It is now obvious that the route satisfies the condition, after we transformed the representation into the abstraction.

Now, we have to show this requirement on \texttt{train\_1} and \texttt{train\_1\_flatten}. Based on the way abstraction is defined, we already \emph{know} that our requirement will hold as our gluing invariant does not strengthen any guard. 
To show this requirement on \texttt{train\_1}, we need to translate \texttt{rs} into an expression in \texttt{train\_1}. 
In our situation, this poses a challenge, as we have three gluing invariants which we describe one route status each. 
We transform the VO with the help of the glueing invariants. 
For example, \texttt{rs(R8)} is translated as: \newline

$a = \newline (\{1 \mapsto (\{x | x \notin (resrt \cup frm) \} \ (ROUTES \setminus \{R8\}))  \} \cup \newline \{2 \mapsto ((resrt \setminus frm) \ (ROUTES \setminus \{R8\}))  \} \cup \newline \{3 \mapsto (frm \setminus (ROUTES \setminus \{R8\}))\})~[\{\{R8\}\}])$ \newline

\begin{figure}
    \centering
    \includegraphics[scale=0.45]{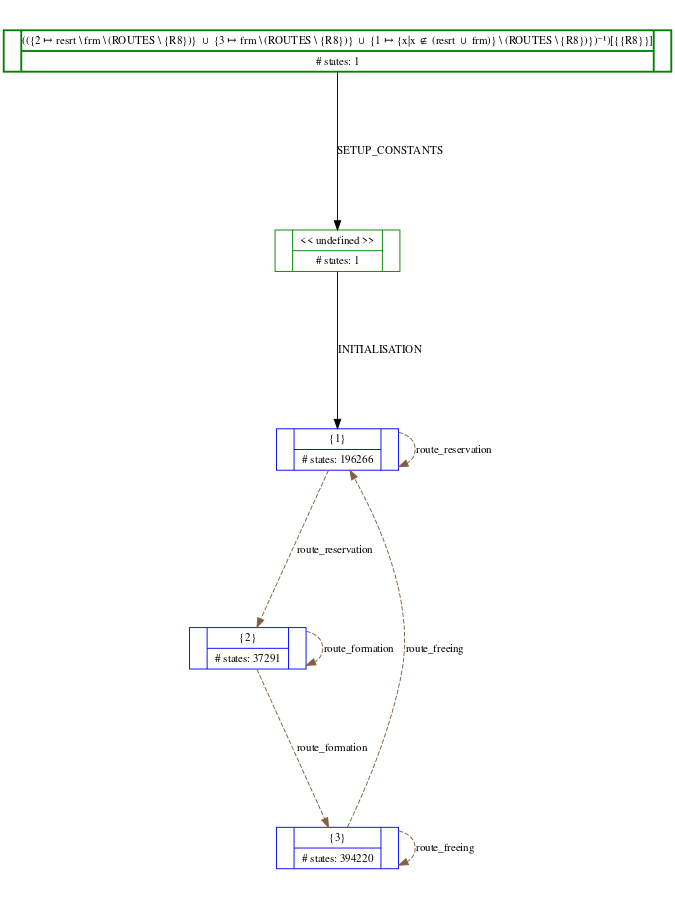}% Huch
    \caption{Projection of \texttt{train\_1}'s State Space projected on Route 8}
    \label{fig:concretePrj}
\end{figure}

This expression re-creates the route status on \texttt{train\_1}. 
1 is the status for free, 
2 for reserved and 
3 for formed. 
The expression in the end is for pretty printing the projection --- due to the nature of ProB's graph visualization, we had to insert maplet operators to separate the different sets.
The last command removes the maplets. 
VO3 is therefore refined to work on the original. We call this refinement VO3.1:
\begin{itemize}
    %\item VO3.1: MC;SPRJ($(\{1 \mapsto (\{x | x \notin (resrt \cup frm) \} \ (ROUTES \setminus \{R8\}))  \} \cup \{2 \mapsto ((resrt \setminus frm) \ (ROUTES \setminus \{R8\}))  \} \cup \{3 \mapsto (frm \setminus (ROUTES \setminus \{R8\}))\})~[\{\{R8\}\}])$, ...
    \item VO3.1: MC3.1(FIN);(SPRJ3.1 \& \ldots)
    \item MC3.1/\texttt{train\_1}/MC(FIN)
    \item SPRJ3.1/\texttt{train\_1}/SPRJ: $(\{1 \mapsto (\{x | x \notin (resrt \cup frm) \} \ (ROUTES \setminus \{R8\}))  \} \cup \{2 \mapsto ((resrt \setminus frm) \ (ROUTES \setminus \{R8\}))  \} \cup \{3 \mapsto (frm \setminus (ROUTES \setminus \{R8\}))\})~[\{\{R8\}\}])$
    \item \ldots
\end{itemize}
The result for route 8 is visible in \cref{fig:concretePrj}. The main difference is now that the line between states is doted. 
In the abstraction, the lines are not dotted which means that the states are direct neighbors in the model, i.e. that reserved is the direct succeeding state of free. 
In \texttt{train\_1}, between the actual reserving and the reserved status, there are other events executed, e.g., \textbf{point\_positioning}.

\newpage

\listoffigures
\listoftables

\addcontentsline{toc}{section}{List of Listings}

\lstlistoflistings 

\newpage

\printbibliography

\end{document}